\documentclass[twocolumn,prl,aps,floatfix,superscriptaddress,amssymb]{revtex4}
\usepackage{epsfig}

\begin{document}

\title{Zero-energy states in triangular and trapezoidal graphene structures}

\author{P.~Potasz}
\affiliation{Institute for Microstructural Sciences, National Research Council of Canada
, Ottawa, Canada}
\affiliation{Institute of Physics, Wroclaw University of Technology, Wroclaw, Poland}

\author{A.~D.~G\"u\c{c}l\"u}
\affiliation{Institute for Microstructural Sciences, National Research Council of Canada
, Ottawa, Canada}

\author{P.~Hawrylak}
\affiliation{Institute for Microstructural Sciences, National Research Council of Canada
, Ottawa, Canada}

\date{\today}

\begin{abstract}
We derive analytical solutions for the zero-energy states of degenerate shell obtained as a singular eigenevalue problem found in tight-binding (TB) Hamiltonian of triangular graphene quantum dots with zigzag edges. These analytical solutions are in agreement with previous TB and density functional theory (DFT) results for small graphene triangles and extend to arbitrary size. We also generalize these solutions to trapezoidal structure which allow us to study bowtie graphene devices. 
\end{abstract}

\maketitle

Low dimensional graphene nanostructures are promising candidates as building
blocks for future nanoelectronic applications due to their band gaps and magnetic properties tunable with size and shape \cite{NGM+04,NGM+05,ZTS+05,RTB+04,GN+05,NGP+09}. Remarkable progress has
been made in cutting graphene sheets into nanostructures with desired shape and size, significantly
influencing their properties \cite{GN+05,NGP+09,CMS+07}. In
particular, the existence of a band of 
degenerate states near Fermi level localized at the edges in zigzag ribbons \cite{NFD+96,Eza09,SCL+10} 
and triangular dots \cite{YNW+11,Eza07,Eza08,FP07,AHM08,WMK08,GPH+02} was 
predicted by tight-binding model and confirmed by density functional theory
calculations. These zero-energy edge states play important role due to their
large contribution to the density of states \cite{YNW+11,FP07,WSS+08}. In triangular graphene quantum dots, numerical results show that the degeneracy of the band of zero-energy states is proportional to the edge size and can be made macroscopic. This opens up the possibility to design a strongly correlated electronic system as a function of filling of the shell, in analogy to the fractional quantum Hall effect \cite{GPH+02}. 

While the existence of zero-energy states was predicted analytically for the
zigzag ribbons \cite{NFD+96}, for triangular structures, the analysis of the zero-energy states was limited to numerical techniques such as tight-binding and density functional theory for specific and small sizes of quantum dots. A size-independent general analytical analysis is therefore desirable. In this work, we present analytical solutions to zero-energy edge states in graphene triangles with zigzag edges. We also show how the results can be generalized to the trapezoidal structures and applied to the bowtie structures \cite{WMK+18}. Our method allows the prediction of the number of zero-energy states as a function of the size in all triangular, trapezoidal and bowtie structures.  

\begin{figure}
\epsfig{file=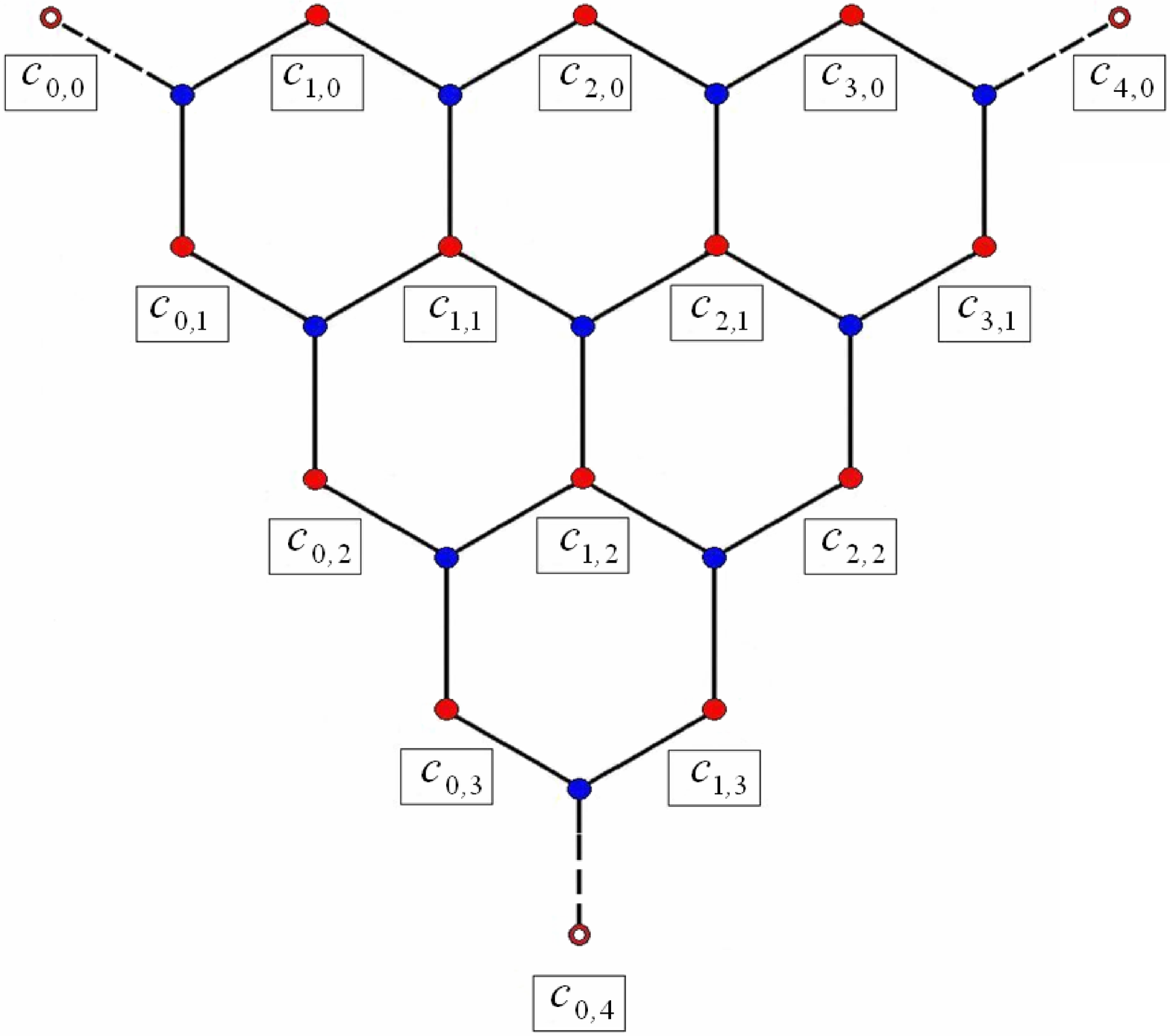,width=3.2in}
\caption{
(Color online) 
Triangular zigzag graphene structure with $N=3$ atoms on the one edge. Under
each $A$--type atom (indicated by red colour) are corresponding
coefficients. Dash lines and open circles indicate auxiliary $A$--type atoms in
the three corners, which will help to introduce three boundary conditions. For
zero-energy states all coefficients can be expressed as superpositions of
coefficients corresponding to atoms from the one edge (upper row of atoms in our case).}\label{fig:trianred}
\end{figure}

Our starting point is the nearest-neighbour tight binding model. It has been successfully used to describe graphene lattice \cite{Wallace} and applied to other graphene materials such as nanotubes, nanoribbons and quantum dots \cite{NFD+96,Eza09,YNW+11,Eza07,Eza08,FP07,AHM08,SDD98}. 
The Hamiltonian is written as 
\begin{eqnarray}
\nonumber
&H&= t\sum_{\left\langle i,j\right\rangle}a^\dagger_{i}a_{j},       
\label{hamiltonian}
\end{eqnarray}
where $t$ is hopping integral, $a^\dagger_{i}$  and $a_{i}$ are creation and annihilation operators
on a site $i$ respectively, and $\left\langle i,j\right\rangle$ indicate summation over nearest-neighbours. It
is important to distinguish between two types of atoms which appear in the
unit cell of the honeycomb lattice of graphene sheet. For triangular structures, these atoms form two
non-equivalent sublattices ($A$ and $B$) and they are indicated by red and blue
circles of the graphene triangle in Fig. \ref{fig:trianred}. 
Our goal is to find zero-energy solutions to the singular eigenvalue problem,
\begin{eqnarray}
\nonumber
H\Psi=0.
\label{singular}
\end{eqnarray} 
In this case there is no coupling between two sublattices and the solutions can be written separately
for $A$--type and $B$--type atoms as $\Psi^\mu=\sum c_{i}\phi_{i}^\mu$ with $\mu=A,B$. The coefficients $c_{i}$ obey 
\begin{eqnarray}
\sum_{\left\langle i,j\right\rangle}c_{i}=0,     
\label{condition}
\end{eqnarray}
where the summation is over $i$-th nearest neighbours of an atom $j$. In other
words, the sum of coefficients around each site must vanish \cite{NFD+96}.
Let's first focus on the sublattice labelled by $A$, represented by red atoms in Fig. \ref{fig:trianred}. We
label each atom by two integer numbers $n$ and $m$ (with $0<n,m<N+1$, where N is the number of $A$--type atoms on the one edge). The dash lines and
open circles indicate auxiliary atoms which will later help to introduce
boundary conditions. We will now show that coefficients $c_{n,m}$ for all atoms in the
triangle can be expressed as a linear combination of coefficients
corresponding to atoms on one edge, i.e. $c_{n,0}$. Starting from the first row and
using Eq. (\ref{condition}), we can obtain all coefficients corresponding to atoms in the
second row. For the first two coefficients from the left we obtain $c_{0,1}=-(c_{0,0}+c_{1,0})$ and $c_{1,1}=-(c_{1,0}+c_{2,0})$. These coefficients are just equal to the sum of two upper lying coefficients with the minus sign. In analogy, we can write expressions for all coefficients in the second row.
In the next step, coefficients in the third row are expressed as a sum of two coefficients in the second row. For first coefficient from the left in the third row we obtain $c_{0,2}=-(c_{0,0}+c_{1,1})=(c_{0,0}+2c_{1,0}+c_{2,0})$. The second and third ones will have similar form. By going down rows one by one, we can obtain all coefficients in the structure regardless of the size of the triangle. Similar to the construction of Pascale triangle \cite{Pascal}, these coefficients can be written in a suitable form using binomial coefficients
\begin{eqnarray}
c_{n,m}=(-1)^m\sum_{k=0}^m\left(\begin{array}{c}
m\\
k\end{array}\right) c_{n+k,0}.
\label{coeff}
\end{eqnarray} 
Here, it is important to emphasize that the only unknown are the $N+2$ coefficients ($c_{n,0}$'s) from the first row; the rest are expressed as their superpositions, as it is seen from Eq. (\ref{coeff}). In addition, we must use the boundary conditions: the construction of the triangle requires vanishing of the coefficients corresponding to auxiliary atoms in each corner (Fig. 1). This gives three boundary conditions ($c_{0,0}=c_{N,0}=c_{0,N}=0$), reducing the number of independent coefficients to $N-1$.

The same analysis can be done for $B$--type atoms indicated by blue colour. In
this case, it is convenient to include some of boundary conditions at the
beginning as shown in Fig. 2, where we only keep coefficients belonging to
auxiliary atoms on the right edge.
\begin{figure}[htbp]
\epsfig{file=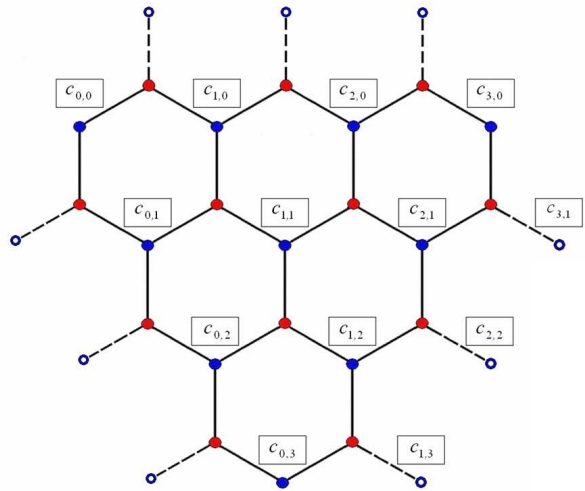,width=3.2in}
\caption{
(Color online) 
Triangular zigzag graphene structure from Fig. 1. Under each $B$--type atom
(indicated by blue colour) are corresponding coefficients. For convenience, we
only left coefficients corresponding to auxiliary $B$--type atoms on the right
edge. For zero energy states coefficient from upper left corner ($c_{0,0}$) determine all
other coefficients in the structure. Introducing three boundary conditions
from auxiliary atoms we obtain only trivial solution; zero-energy states
consist of only $A$--type atoms.}\label{fig:trianblue}  
\end{figure} 
As a consequence, the coefficient $c_{0,0}$
determines all other coefficients in the triangle. Since there are three auxiliary
atoms (equivalently three boundary conditions) but only one independent coefficient,
we can not obtain any nontrivial solution. Hence, zero-energy states can only consist
of coefficients of one type atoms -- these lying on the edges. Now we can write
general form for the eigenvectors for zero-energy states in the triangle 
\begin{eqnarray}
\Psi=\sum_{n=0}^{N+1}\sum_{m=0}^{N+1-n}\left[(-1)^m\sum_{k=0}^m\left(\begin{array}{c}
m\\
k\end{array}\right) c_{n+k,0}\right]\phi_{n,m}^A,       
\label{triangle}
\end{eqnarray}
where $N$ is the number of atoms on the one edge and $\phi_{n,m}^A$ is $p_{z}$ orbital
on $A$--type site ($n,m$). In this expression the only $N-1$ coefficients corresponding to atoms
from the first row are independent. Thus, we can construct $N-1$ linear
independent eigenvectors which span the subspace with zero-energy states. This
is in agreement with Ref. \cite{FP07} -- the number of zero-energy states in the
triangle is $N-1$, where $N$ is the number of atoms on one edge.
  
Using the Eq. (\ref{triangle}) we can then construct an orthonormal basis for
zero-energy states. First, we make a choice for the $N-1$ independent 
coefficients $c_{n,0}$, from which we obtain $N-1$ linear independent vectors, for
instance by choosing only one nonzero coefficient for all $N-1$ collections,
different one for each eigenvector. Resulting eigenvectors can then be
orthogonalized using standard Gram--Schmidt process. The last step is the
normalisation $K_{norm}$ of the eigenvectors, using expression
\begin{eqnarray}
\nonumber
K_{norm}=\sum_{n=0}^{N+1}\sum_{m=0}^{N+1-n}\left|\sum_{k=0}^m
\left(\begin{array}{c}
m\\
k\end{array}\right)
c_{n+k,0}\right|^2.       
\label{norm}
\end{eqnarray}
 
The method for obtaining zero-energy eigenfunction coefficients for the
triangular structures can also be applied to trapezoidal structures (inset of
Fig. 3(b)).
\begin{figure}
\epsfig{file=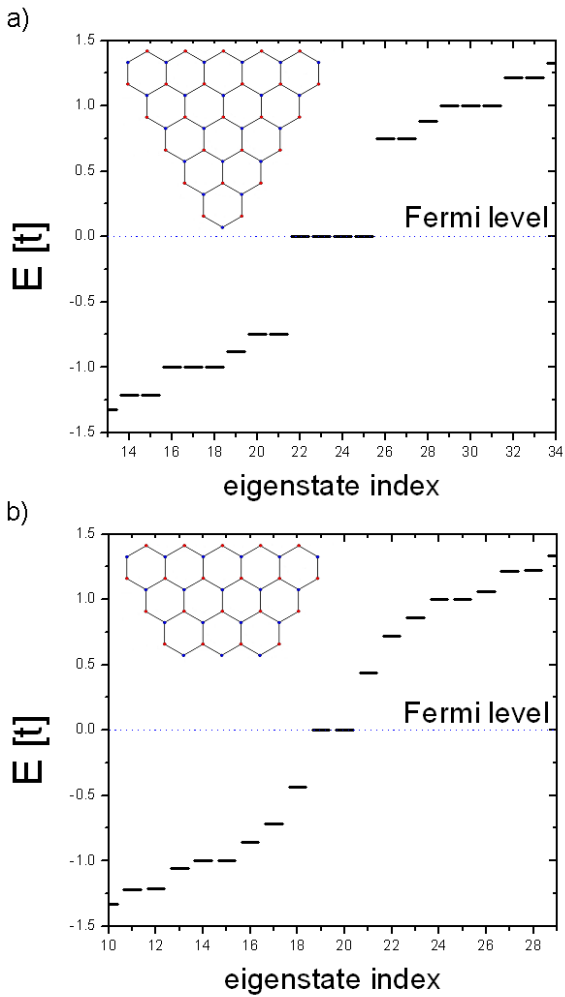,width=3.2in}
\caption{
(Color online) 
Single particle spectrum from tight-binding calculations for a) 46 atoms
triangle and b) 38 atoms trapezoid (with $N_{row}=4$). There are four zero-energy states
in the triangle (number of atoms on the one edge $N=5$) and two zero-energy states in
the trapezoid in agreement with our analysis. Changing number of rows in the
trapezoid we can control number of zero-energy states.}\label{fig:trapeze}  
\end{figure} 
As explained above, the value of the coefficients for atoms in a
given row is sufficient to determine the coefficients for atoms in the lower
lying row. If we stop this process of going down the ladder one by one at any
row, we then obtain a trapezoidal structure. The Eq. (\ref{triangle}) takes the following form 
\begin{eqnarray}
\Psi=\sum_{n=0}^{N+1}\sum_{m=0}^{M}\left[(-1)^m\sum_{k=0}^m\left(\begin{array}{c}
m\\
k\end{array}\right) c_{n+k,0}\right]\phi_{n,m}^A,       
\label{trapez}
\end{eqnarray}
where $M=min(N+1-n,N_{row}-1)$ and $N_{row}$ is the number of rows in the structure
(see Fig. \ref{fig:trapeze}(b)). In this case the last row contains $N-N_{row}+2$ auxiliary atoms which 
increases the number of boundary conditions. The number of zero-energy states is
then given by $N_{row}-2$ (for $N_{row}>1$). Here we note that similar to the triangle, 
zero-energy states consist of only one type of atoms; the only difference is increased
number of boundary conditions. In Fig. \ref{fig:trapeze}(a) we show tight-binding single
particle states for triangle with $N=5$ atoms on one edge. As expected, there
are four zero-energy states. For comparison, in Fig. \ref{fig:trapeze}(b) we show single
particle states for trapezoid with the same number of atoms in a first
row. Here, there are only two zero-energy states in agreement with our
analysis -- increasing number of boundary conditions decrease
number of zero-energy states. We note that the structure which
consists of only two rows (the single chain of benzene rings, called acene)
does not have zero-energy states while the triangular structure with $N$ atoms
on the one edge has maximal number of zero-energy states equal to $N-1$. All
intermediate structures (trapezoidal structures) have number of zero-energy
states in the range between $1$ and $N-2$, depending on the number of rows. 

\begin{figure}
\epsfig{file=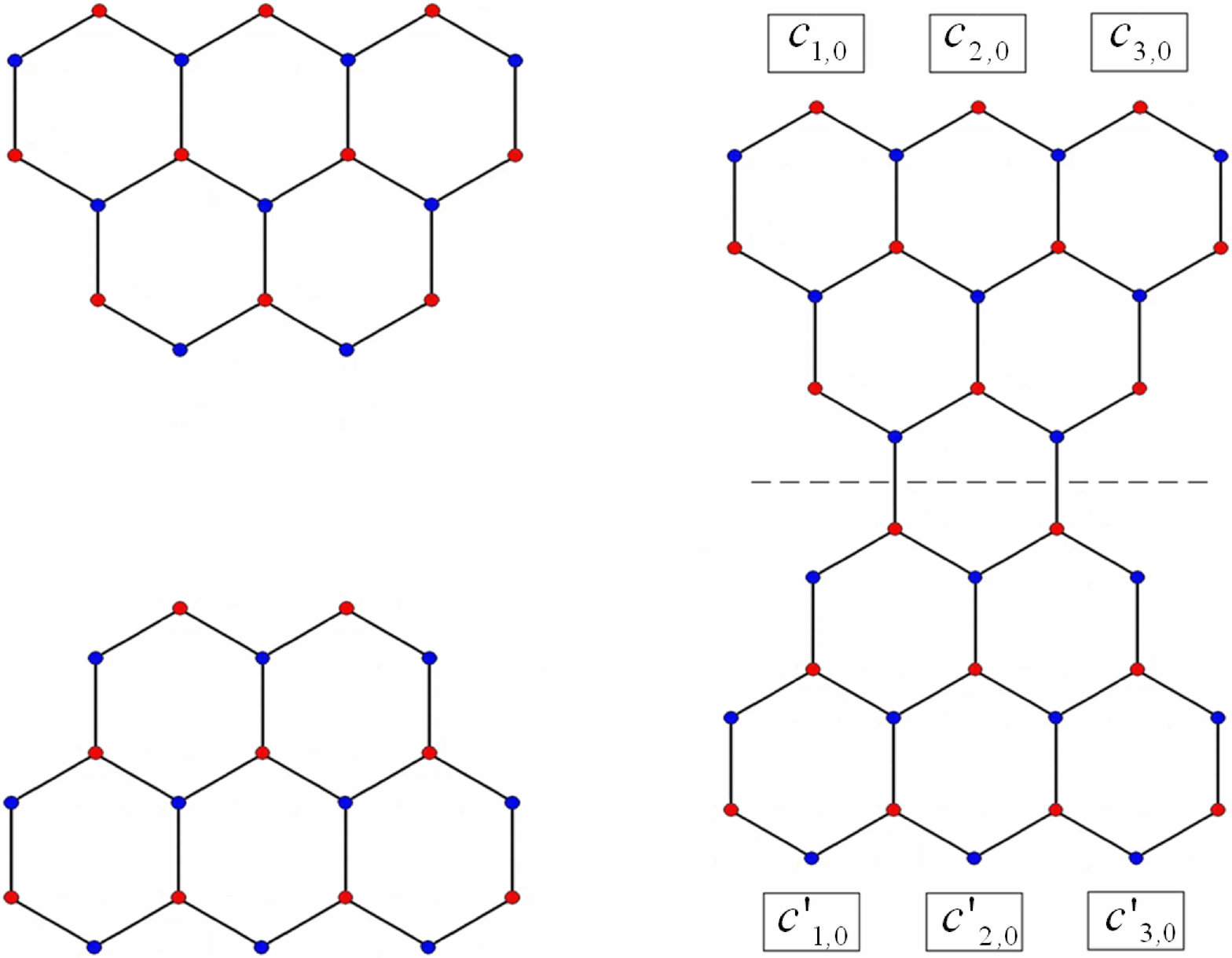,width=3.2in}
\caption{
(Color online) 
Two trapezoids (left) and bowtie structure (right). Each of two trapezoids has
one zero-energy state, consisting of only $A$--type atoms (indicated by red
colour) for the upper trapezoid, and consisting of only $B$--type atoms
(indicated by blue colour) for the lower trapezoid. Connecting these two systems
does not affect the zero-energy solutions since coefficients belonging to
connecting atoms are zeros (the four nearest atoms to the dash line). The
bowtie structure on the right has two zero-energy states: one which completely
lies in upper part and consists of $A$--type atoms, and second one lies in lower
part and consists of $B$--type atoms.}\label{fig:trapbowtie}   
\end{figure}

Finally we note that the solutions of Eq. (\ref{trapez}) can also be applied to bowtie
structures \cite{WMK+18}. These can be treated as two trapezoidal structures connected
by their shorter base, shown on Fig. \ref{fig:trapbowtie}.
It is important to emphasize that the upper trapezoid has one zero-energy state 
which consists of $A$--type atoms (red atoms) while lower trapezoid has one zero-energy state
which consists of $B$--type atoms (blue atoms). Connecting these two systems does not
affect the zero-energy solutions since coefficients belonging to connecting
atoms are zeros. Using zero-energy eigenvectors for trapezoids, Eq. (\ref{trapez}), we
obtain expressions for two groups of zero-energy states in the bowtie structures   
\begin{eqnarray}
\Psi=\sum_{n=0}^{N+1}\sum_{m=0}^{M}\left[(-1)^m\sum_{k=0}^m\left(\begin{array}{c}
m\\
k\end{array}\right) c_{n+k,0}\right]\phi_{n,m}^A, 
\label{bowtie}  
\end{eqnarray}
for upper trapezoid, where $A$ indicate $A$--type atoms from upper part and 
\begin{eqnarray}    
\Psi=\sum_{n=0}^{N'+1}\sum_{m=0}^{M'}\left[(-1)^m\sum_{k=0}^m\left(\begin{array}{c}
m\\
k\end{array}\right) c'_{n+k,0}\right]\phi_{n,m}^{B'},       
\label{bowtie2}
\end{eqnarray}
for lower one, where $B'$ indicate $B$--type atoms from lower part.
Two parts of the bowtie structure are separated by the dash
line in Fig. \ref{fig:trapbowtie}. Coefficients $c_{n,0}$($c'_{n,0}$) correspond to $N$ $A$--type ($N'$�$B$--type) atoms from
the highest (lowest) row in the bowtie structure from Fig. \ref{fig:trapbowtie}. Note that it is
possible to use Eqs. (\ref{bowtie}) and (\ref{bowtie2}) to asymmetric bowtie structures consisting of two
different trapezoids ($N\neq N'$).   

In summary, we derived here analytical expression for zero-energy states in triangular and trapezoidal graphene quantum dot structures. Our method allows prediction of the number of zero-energy states in quantum dots of arbitrary size which can be understood in terms of a competition between the number of independent coefficients and the number of auxiliary atoms (the number of boundary conditions). We also showed that the number of zero-energy states can be controlled by changing the number of rows in the trapezoidal structures but does not depend on the number of atoms in the base of the trapezoid. Finally, we applied our results to bowtie structures and showed that two independent groups of zero-energy states coexist in these systems.
  

{\it Acknowledgement}. The authors thank NRC-CNRS CRP, Canadian Institute for
Advanced Research, Institute for Microstructural Sciences, QuantumWorks
and Polish MNiSW, Grant No. N202-071-32/1513 for support.

\vspace*{-0.22in}


\end{document}